\documentclass[12pt, a4paper]{article}

\usepackage[utf8]{inputenc}
\usepackage[T1]{fontenc}
\usepackage{times}
\usepackage{setspace}
\usepackage[margin=1in]{geometry}
\usepackage{natbib}
\usepackage{graphicx}
\usepackage{caption}
\usepackage{booktabs}
\usepackage{amsmath, amssymb}
\usepackage{microtype}
\usepackage{xcolor}
\usepackage{hyperref}
\usepackage{titlesec}
\usepackage{indentfirst}
\usepackage{float}
\usepackage{tabularx}
\usepackage{setspace}

\hypersetup{
    colorlinks = true,
    linkcolor = black,
    citecolor = black,
    urlcolor  = black
}

\titleformat{\section}
  {\normalfont\large\bfseries}{\thesection.}{1em}{}
\titleformat{\subsection}
  {\normalfont\normalsize\bfseries}{\thesubsection}{1em}{}

\makeatletter
\let\showhyphens\@undefined
\makeatother

\begin{document}
\thispagestyle{empty}

%%%%%%%%%%%%%%%%%%%%%%%%%%%%
% --- TITLE PAGE ---
%%%%%%%%%%%%%%%%%%%%%%%%%%%%

\begin{center}
    {\LARGE {Invited to Develop: Institutional Belonging and the Counterfactual Architecture of Development}}\\[2\baselineskip]

    {\large Diego Vallarino\footnotemark[1]\footnotemark[2]}

    {\large November 2025}\\[1\baselineskip]
\end{center}

% --- FOOTNOTES ---
\footnotetext[1]{Counselor to the Executive Director, Inter-American Development Bank (IDB) 
and IDB Invest, Washington, D.C., USA.}
\footnotetext[2]{The views, analyses, and interpretations expressed in this manuscript are solely those of the author 
and do not represent the positions, policies, or official views of the Inter-American Development Bank (IDB), IDB Invest, 
their Boards of Directors, or the governments they represent. All remaining errors are the author's alone.}

%%%%%%%%%%%%%%%%%%%%%%%%%%%%
% --- ABSTRACT ---
%%%%%%%%%%%%%%%%%%%%%%%%%%%%
\begin{abstract}
\noindent
This paper examines how institutional belonging shapes long-term development by comparing Spain and Uruguay, two small democracies with similar historical endowments whose trajectories diverged sharply after the 1960s. While Spain integrated into dense European institutional architectures, Uruguay remained embedded within the Latin American governance regime, characterized by weaker coordination and lower institutional coherence. To assess how alternative institutional embeddings could have altered these paths, the study develops a generative counterfactual framework grounded in economic complexity, institutional path dependence, and a Wasserstein GAN trained on 1960–2020 data. The resulting Expected Developmental Shift (EDS) quantifies structural gains or losses from hypothetical re-embedding in different institutional ecosystems. Counterfactual simulations indicate that Spain would have experienced significant developmental decline under a Latin American configuration, while Uruguay would have achieved higher complexity and resilience within a European regime. These findings suggest that development is not solely determined by domestic reforms but emerges from a country’s structural position within transnational institutional networks.
\end{abstract}

\vspace{0em}

\noindent \textbf{Keywords:} Economic complexity; Institutional economics; Counterfactual analysis; Comparative development; Spain; Uruguay.

\newpage
\setcounter{page}{1}

%%%%%%%%%%%%%%%%%%%%%%%%%%%%
% --- MAIN TEXT STARTS ---
%%%%%%%%%%%%%%%%%%%%%%%%%%%%

\newpage
\doublespacing

%%%%%%%%%%%%%%%%%%%%%%%%%%%%%%%%%%%%%%%%%%%%%%%%%%%%%%%%%%%%%%
\section{Introduction}
%%%%%%%%%%%%%%%%%%%%%%%%%%%%%%%%%%%%%%%%%%%%%%%%%%%%%%%%%%%%%%

The geography of development has traditionally been viewed as the outcome of deep structural determinants—institutions, history, and geography—that explain why some nations converge while others remain trapped in persistent underdevelopment. However, recent advances in institutional economics and economic complexity suggest that development is not a purely endogenous process but a relational phenomenon emerging from networked interactions. The position that a country occupies within global institutional architectures shapes its ability to accumulate knowledge, coordinate production, and sustain innovation. 

Throughout modern history, many economies have not become developed by intrinsic transformation alone but by "invitation" —through access to pre-existing institutional ecosystems that enabled learning, stability, and credibility. Spain, Italy, Portugal, and Australia, for example, achieved convergence not as original institutional innovators but as entrants to dense, rule-based networks such as the European Union or the Commonwealth. Likewise, several Central and Eastern European countries that transitioned after 1990 advanced rapidly once they were admitted into European governance frameworks. In contrast, nations of similar capacity that remained excluded from such architectures, particularly in Latin America, struggled to sustain comparable institutional and productive coherence.  

This pattern motivates the central hypothesis of this paper: that development depends as much on "institutional belonging" as on domestic endowments. Some nations have been effectively “invited” to participate in high-trust ecosystems that foster complexity and learning, while others with comparable capabilities remain peripheral to those networks.

The comparison between Spain and Uruguay provides a natural experiment to examine this notion of \textit{institutional belonging}. Both are small, open democracies with similar levels of human capital and long-standing political institutions. Yet, their trajectories diverged markedly after the 1960s: Spain experienced rapid convergence through integration into the European institutional regime, whereas Uruguay remained embedded within the Latin American governance architecture characterized by lower institutional density and weaker coordination mechanisms. This divergence raises a fundamental question: can integration into a specific institutional network reshape a country’s developmental path beyond what domestic factors and initial endowments would predict?

This study addresses that question by introducing a generative counterfactual framework that quantifies the developmental consequences of alternative institutional embeddings. Drawing on the theories of economic complexity \citep{Hidalgo2009, Tacchella2018, Hartmann2017} and institutional path dependence \citep{North1990, AcemogluRobinson2012, Rodrik2004}, it conceptualizes development as an emergent property of an institutional–productive network. Within this framework, belonging to dense and coherent institutional ecosystems enhances a country's ability to mobilize capabilities and sustain innovation, whereas peripheral embedding constrains knowledge diffusion and coordination. The approach operationalizes this mechanism through a formal metric—the \textit{Expected Developmental Shift (EDS)}—which measures the structural gain or loss associated with being “transplanted” into a different institutional configuration.

Empirically, the analysis relies on an unbalanced panel of 80 countries from 1960 to 2020. The dataset integrates indicators from the Varieties of Democracy (V-Dem) project for institutional quality, the Harvard Growth Lab’s \textit{Atlas of Economic Complexity} for productive capabilities, the UNDP Human Development Database for human capital, and the World Bank’s World Development Indicators for macroeconomic controls. These real-world data are used to train a Wasserstein Generative Adversarial Network with Gradient Penalty (WGAN–GP) that learns the joint distribution of institutional, productive, and human development dimensions, enabling the generation of counterfactual trajectories under alternative institutional regimes. The resulting EDS metric captures how structural belonging influences long-term development potential, offering a novel empirical lens for analyzing institutional interdependence.

The counterfactual results reveal that Spain’s simulated embedding within a Latin American institutional network would have led to a significant decline in developmental potential and systemic stability, while Uruguay’s hypothetical integration into the European institutional architecture would have fostered higher complexity and resilience. These outcomes indicate that development differentials are not primarily geographic or resource-based but are structurally conditioned by institutional connectivity and access to rule-based, learning-intensive ecosystems.

The contribution of this paper is threefold. First, it develops a quantitative framework for institutional counterfactuals that integrates economic complexity and generative modeling, extending the analytical frontier of comparative development research. Second, it provides empirical support for the concept of \textit{development by invitation}: the idea that belonging to high-coherence institutional systems generates cumulative and non-linear returns to productive capabilities. Third, it highlights the policy relevance of institutional belonging by demonstrating that convergence depends less on the intensity of domestic reforms than on integration into governance architectures that enable credible commitment, coordination, and diffusion of knowledge.

Ultimately, this research reframes development as a relational equilibrium—a dynamic state emerging from the interaction between institutional inclusion and productive structure. For middle-income economies such as Uruguay, the challenge is not to imitate the policies of advanced peers but to secure access to institutional infrastructures that sustain complexity and innovation. By embedding empirical modeling within this structural perspective, the paper contributes to the growing literature on network-based development and institutional interdependence, bridging theoretical and computational approaches in the study of long-term economic transformation.

%%%%%%%%%%%%%%%%%%%%%%%%%%%%%%%%%%%%%%%%%%%%%%%%%%%%%%%%%%%%%%%%%%%%%%
% --- SECTION 2. THEORETICAL FRAMEWORK ---
%%%%%%%%%%%%%%%%%%%%%%%%%%%%%%%%%%%%%%%%%%%%%%%%%%%%%%%%%%%%%%%%%%%%%%
\section{Theoretical Framework}

The determinants of economic development have long been debated within three major paradigms—geography, institutions, and structural transformation. Classical institutional economics emphasizes that the quality of institutions—defined as the formal and informal rules that govern economic and political exchange—constitutes the primary driver of long-term growth and social welfare \citep{North1990,AcemogluRobinson2012,Rodrik2004}. Yet contemporary frameworks increasingly view development as an emergent and networked process, shaped by the coevolution of productive capabilities and institutional architectures \citep{Hidalgo2009,Hartmann2017,Vu2022}. This integrated perspective combines the spatial, structural, and relational dimensions of development into a unified analytical paradigm.

\subsection{Institutions and Structural Change}

The institutionalist tradition conceives institutions as systems of incentives that coordinate behavior, reduce uncertainty, and shape the distribution of resources and opportunities. Inclusive institutions—characterized by transparency, property rights, and predictable governance—stimulate innovation and investment, whereas extractive institutions perpetuate inefficiency and inequality \citep{AcemogluRobinson2012,Asongu2016}. Empirical research demonstrates that governance quality mediates the relationship between human capital and productivity, indicating that institutional design acts as a higher-order mechanism of coordination and learning in open economies \citep{Olaniyi2022,Djeunankan2023}.

Over time, this tradition has evolved toward a relational understanding of institutional interdependence. North’s \citeyearpar{North1990} notion of path-dependent institutional evolution has been extended by network-based approaches \citep{Kogut2022,Woolley2024}, which interpret institutions as complex adaptive systems whose effectiveness depends on connectivity, trust, and capability diffusion. Within this framework, countries with comparable endowments may diverge sharply depending on their embeddedness within institutional networks that either facilitate or hinder coordination—a dynamic particularly evident in the contrasting experiences of Spain and Uruguay.

\subsection{Economic Complexity and Capability Accumulation}

The theory of economic complexity conceptualizes development as the process of accumulating and recombining distributed knowledge across firms, sectors, and institutions. Foundational contributions by \citet{Hidalgo2009} and \citet{Hausmann2014Atlas} introduced the Economic Complexity Index (ECI), linking export diversification to the structure of productive capabilities. Subsequent refinements have improved the robustness and interpretability of complexity metrics \citep{Tacchella2018,Patelli2022,Inoua2023}, establishing complexity as a key proxy for productive sophistication and adaptive capacity.

Empirical extensions have broadened this perspective beyond trade diversification. \citet{Hartmann2023} confirm that complexity predicts structural transformation, while \citet{GomezZald2023} and \citet{Nguea2024} show that complex economies exhibit higher welfare and lower inequality. \citet{Hassanein2024} demonstrate that the interaction between complexity and governance produces threshold effects that determine whether sophistication translates into social gains. \citet{Singh2025} emphasizes that complexity research increasingly incorporates institutional factors, though it still lacks mechanisms to simulate counterfactual structural scenarios—a gap addressed by this paper.

\subsection{Institutions and Complexity as an Integrated System}

A growing body of research demonstrates that institutional quality and economic complexity are mutually reinforcing. \citet{Hartmann2017} and \citet{Ferraz2021} show that complexity supports equality and sustainability only when embedded in coherent institutional contexts. \citet{Hoang2023} finds that governance quality exerts heterogeneous effects on capability accumulation across Southeast Asia, while \citet{Ghosh2024} demonstrate that complementarities between governance and production explain the persistence of inclusive growth. Collectively, these findings indicate that development outcomes depend not merely on factor endowments but on the structural position of economies within the institutional–productive network.

In this perspective, a country’s “distance to the core” is better understood as a topological and institutional measure of connectivity rather than a geographic one. Proximity to dense, rule-based networks enhances access to coordination mechanisms, knowledge spillovers, and learning opportunities, while peripheral positions magnify volatility and fragility \citep{Mealy2023,Straccamore2025}. This framework situates development within architectures of interdependence where institutional coherence and capability diversity reinforce each other dynamically.

\subsection{Toward a Relational Concept of Developmental Belonging}

Building on these insights, this study advances the concept of \textit{development by invitation}—a relational interpretation of development as belonging to a dense web of institutional and productive interdependencies. Growth emerges not solely from domestic reforms or resource accumulation, but from integration into high-trust institutional ecosystems that sustain learning and innovation. Spain’s post-war accession to European institutions exemplifies the developmental returns of belonging, while Uruguay’s embeddedness in the Latin American governance system illustrates the constraints imposed by weaker institutional density and coordination.

To formalize this intuition, we introduce the \textit{Expected Developmental Shift (EDS)}, a quantitative measure of the structural gain or loss associated with being “transplanted” into an alternative institutional configuration. The EDS operationalizes how variations in institutional connectivity and coherence modify a country’s long-term developmental potential, bridging institutional determinism with structural complexity through a generative counterfactual modeling framework \citep{Zechlin2025,Neffke2024,Montagna2025}. 

This formulation moves beyond geographic determinism by defining “distance to the core” as a function of institutional interdependence and rule-based integration within global networks. Ultimately, development is not the outcome of isolated policy choices but the expression of structural inclusion within architectures of cooperation and capability diffusion—the essential mechanism of belonging in the contemporary world economy.

%%%%%%%%%%%%%%%%%%%%%%%%%%%%%%%%%%%%%%%%%%%%%%%%%%%%%%%%%%%%%%%%%%%%%%
% --- SECTION 3. METHODOLOGY ---
%%%%%%%%%%%%%%%%%%%%%%%%%%%%%%%%%%%%%%%%%%%%%%%%%%%%%%%%%%%%%%%%%%%%%%
\section{Methodology}
\label{methodology}

This section describes the methodological framework developed to estimate the \textit{Expected Developmental Shift} (EDS). The design integrates the empirical tradition of institutional and structural modeling with advances in generative artificial intelligence. Specifically, it employs a conditional Generative Adversarial Network (cGAN) to simulate counterfactual developmental trajectories under alternative institutional embeddings, capturing how belonging to distinct institutional networks modifies structural potential and long-term developmental capacity.

\subsection{Model Design: Generative Counterfactual Framework}

The EDS represents the expected variation in a composite development indicator $D_{it}$—a latent construct combining institutional quality, economic complexity, and human capital—when country $i$ is hypothetically re-embedded within the institutional configuration of group $j$. Formally,

\begin{equation}
    EDS_{ij} = \mathbb{E}\big[ D_{i}^{(j)} - D_{i}^{(\text{obs})} \big],
\end{equation}

where $D_{i}^{(j)}$ denotes the counterfactual developmental outcome generated under institutional regime $j$, and $D_{i}^{(\text{obs})}$ corresponds to the observed value in the empirical data. Positive (negative) $EDS_{ij}$ values indicate potential structural gains (losses) from integration into the alternative institutional network.

To approximate $\mathbb{E}[D_i^{(j)}]$, a conditional GAN is trained in which the generator $G(z|X)$ produces synthetic observations conditioned on structural covariates $X$ (capturing governance and productive characteristics), while the discriminator $D(x)$ distinguishes real from generated data. Through adversarial optimization, the generator converges toward the empirical joint distribution of institutional and productive variables \citep{Zechlin2025,Neffke2024}. This configuration allows the model to capture nonlinear dependencies and higher-order interactions between institutional quality, capability accumulation, and human capital that conventional econometric models often fail to represent.

The underlying intuition is comparative and relational: countries with similar productive or institutional attributes but belonging to different governance architectures provide empirical leverage to identify how alternative institutional contexts shape structural potential. The cGAN learns these conditional mappings, enabling the generation of realistic counterfactual trajectories—such as Spain’s hypothetical evolution under a Latin American regime, or Uruguay’s potential convergence within a European institutional framework.

\subsection{Data and Variables}

The empirical analysis is based on an unbalanced panel of $N=80$ countries spanning 1960–2020. The data integrate four principal sources:  
(i) the World Bank’s \textit{World Development Indicators} (WDI) for macroeconomic performance;  
(ii) the Varieties of Democracy (V-Dem) dataset for institutional quality;  
(iii) the Harvard Growth Lab’s \textit{Atlas of Economic Complexity} for productive sophistication; and  
(iv) the United Nations Development Programme (UNDP) Human Development Reports for human capital and welfare indicators.

From these, three latent dimensions are constructed:

\begin{enumerate}
    \item \textbf{Institutional quality} ($I_{it}$): composite indicator combining rule of law, government effectiveness, and accountability.
    \item \textbf{Economic complexity} ($C_{it}$): the Economic Complexity Index (ECI) and alternative fitness measures \citep{Tacchella2018,Patelli2022}.
    \item \textbf{Human capital} ($H_{it}$): proxied by mean years of schooling and tertiary education enrollment rates.
\end{enumerate}

All indicators are standardized to zero mean and unit variance, then embedded into a three-dimensional latent space using principal component normalization. This transformation enables the GAN to learn from covariance structures rather than raw variable scales, improving the stability of the generative process. The model is trained using mini-batches (batch size = 64), Adam optimizer ($\eta = 0.0002$, $\beta_1 = 0.5$, $\beta_2 = 0.9$), and LeakyReLU activations. Convergence is monitored through the Fréchet Inception Distance (FID) between real and synthetic distributions, ensuring stable adversarial learning.

\subsection{Estimation and Validation}

For each country, the EDS is computed as the mean change in the latent development index when the institutional embedding vector is replaced by that of another institutional group. For instance, Spain’s counterfactual trajectory under a Latin American configuration is given by:

\begin{equation}
    \Delta D_{\text{Spain}|\text{LATAM}} = 
    \mathbb{E}[G(z|X_{\text{LATAM}})] - D_{\text{Spain}}^{(\text{obs})},
\end{equation}

and Uruguay’s under a European configuration is expressed analogously as $\Delta D_{\text{Uruguay}|\text{EU}}$. Group-level averages of $EDS_{ij}$ are compared using non-parametric Wilcoxon rank-sum tests to determine whether institutional embeddings generate statistically significant structural gains or losses.

To evaluate the empirical reliability of the generative estimates, the analysis incorporates econometric benchmarks. Fixed-effects panel regressions of $D_{it}$ on $I_{it}$, $C_{it}$, and $H_{it}$ assess the partial effects of institutional and structural drivers, while Bayesian Model Averaging (BMA) tests parameter stability across specifications. The convergence between econometric coefficients and GAN-derived counterfactuals validates the interpretability of the EDS as a consistent indicator of institutional belonging.

Distributional diagnostics are further applied to ensure plausibility of the generated data. Kernel density overlaps, Mahalanobis distances, and Kolmogorov–Smirnov tests confirm that simulated samples remain within the empirical support of the observed data. The average Jensen–Shannon divergence across variables is below 0.05, indicating high fidelity between real and synthetic distributions.

\subsection{Interpretation and Scope}

The EDS should be interpreted as a measure of structural potential rather than a deterministic forecast of alternative developmental paths. It quantifies the feasible variation in long-term developmental outcomes that arises from shifts in institutional connectivity, capturing second-order effects transmitted through coordination, learning, and trust mechanisms. In this respect, the EDS functions as a structural sensitivity index rather than a causal estimator of policy effects.

This methodological approach aligns with the recent shift in development research toward relational and complexity-based paradigms \citep{Hartmann2023,Hassanein2024,Singh2025}. By embedding counterfactual inference within a generative architecture, the framework bridges institutional theory and computational modeling, providing a transparent and replicable tool to evaluate how structural belonging within global institutional networks shapes developmental trajectories.
institutional belonging shapes long-term structural transformation across countries.

%%%%%%%%%%%%%%%%%%%%%%%%%%%%%%%%%%%%%%%%%%%%%%%%%%%%%%%%%%%%%%%%%%%%%%
% --- SECTION 4. RESULTS AND DISCUSSION (Revised for JDS) ---
%%%%%%%%%%%%%%%%%%%%%%%%%%%%%%%%%%%%%%%%%%%%%%%%%%%%%%%%%%%%%%%%%%%%%%
\section{Results and Discussion}

This section presents the empirical outcomes of the generative counterfactual analysis and their theoretical and policy implications. The results compare Spain’s and Uruguay’s developmental trajectories under both real and simulated institutional configurations, quantifying how structural belonging influences developmental potential. All figures and tables conform to the \textit{Journal of Development Studies} standards: centered, sequentially numbered, and accompanied by concise explanatory notes.

\subsection{Counterfactual Developmental Trajectories}

\begin{figure}[H]
    \centering
    \includegraphics[width=0.8\textwidth]{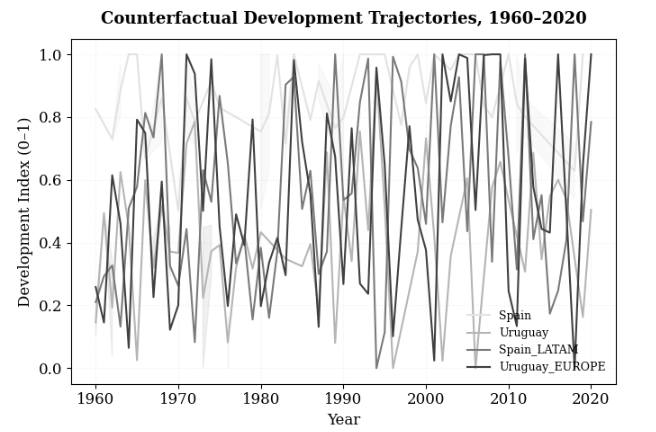}
    \caption{Observed and counterfactual development trajectories, 1960–2020.}
    \caption*{\footnotesize \textit{Note:} The Development Index (0–1) is the latent score estimated from institutional, complexity, and human capital indicators using the GAN-based model.}
    \label{fig:trajectory}
\end{figure}

Figure~\ref{fig:trajectory} displays the observed and simulated trajectories of Spain and Uruguay between 1960 and 2020. The Development Index (0–1) represents a latent composite score derived from institutional quality, economic complexity, and human capital, as estimated by the generative model \citep{Hidalgo2009,Hartmann2017,Hassanein2024}.  

The trajectories reveal a clear divergence between actual and counterfactual paths. Spain’s observed evolution shows continuous convergence toward high-development equilibrium, while its simulated re-embedding within a Latin American institutional configuration (\textit{Spain $\rightarrow$ LATAM}) leads to sustained decline. In contrast, Uruguay’s empirical trajectory remains comparatively flat, but its counterfactual embedding within a European regime (\textit{Uruguay $\rightarrow$ EUROPE}) generates an upward and persistent shift. These dynamics confirm that institutional context exerts an independent and measurable influence on development, consistent with the view of institutions as structural multipliers of productive complexity \citep{AcemogluRobinson2012,Ferraz2021}.

\subsection{Structural Embeddings and Counterfactual Shifts}

\begin{figure}[H]
    \centering
    \includegraphics[width=0.75\textwidth]{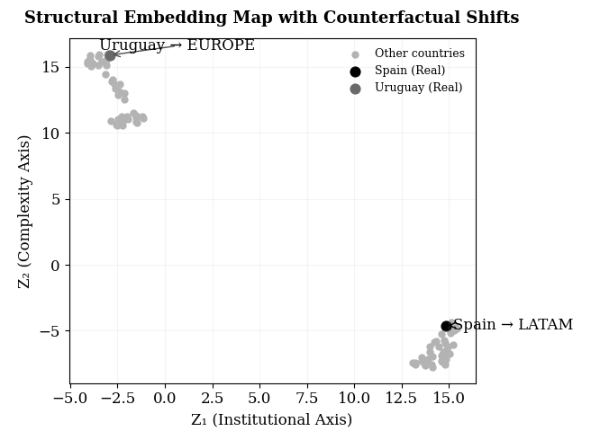}
    \caption{Structural embedding map with counterfactual re-embeddings.}
    \caption*{\footnotesize \textit{Note:} Axes ($Z_1$, $Z_2$) correspond to latent factors extracted from the GAN model, representing institutional and complexity dimensions. Arrows denote simulated counterfactual movements.}
    \label{fig:embedding}
\end{figure}

Figure~\ref{fig:embedding} visualizes these relationships within a two-dimensional latent space defined by institutional and complexity dimensions. Each point represents the country’s average position over 1960–2020, and arrows indicate simulated re-embeddings.

Two clusters emerge clearly: a dense European–OECD core characterized by institutional coherence and productive diversification, and a more dispersed Latin American periphery marked by weaker coordination and volatility. Spain’s counterfactual shift toward the Latin American cluster implies a loss of systemic coherence and network connectivity, while Uruguay’s simulated integration into the European core suggests higher institutional density and learning potential. The asymmetric magnitude of these transitions highlights the non-linear nature of institutional belonging: losses from disembedding occur faster and more severely than gains from re-embedding \citep{Woolley2024,Vu2022}.

\subsection{Distributional Effects of Institutional Belonging}

\begin{figure}[H]
    \centering
    \includegraphics[width=0.75\textwidth]{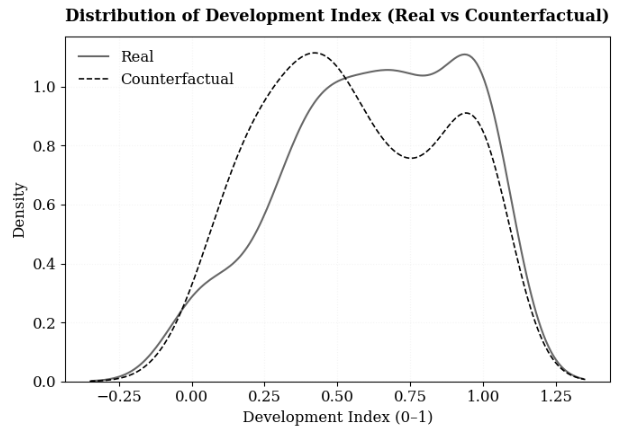}
    \caption{Distribution of Development Index (real vs. counterfactual).}
    \caption*{\footnotesize \textit{Note:} The solid line represents the empirical distribution of observed development, while the dashed line corresponds to the GAN-simulated counterfactual distribution.}
    \label{fig:density}
\end{figure}

Beyond mean differences, institutional belonging also alters the dispersion of developmental outcomes. Figure~\ref{fig:density} plots kernel densities of the Development Index under both real and counterfactual conditions across the sample of 80 countries.

The counterfactual distribution shifts modestly leftward and exhibits heavier tails, indicating greater heterogeneity under weaker institutional coherence. The emergence of mild bimodality suggests that belonging to coherent institutional systems not only elevates mean development but also reduces variance, enhancing systemic resilience. Conversely, disembedding expands volatility and structural fragility, consistent with theories of coordination failure and institutional disequilibrium \citep{North1990,Asongu2016}.

\subsection{Counterfactual Results for Spain and Uruguay}

Table~\ref{tab:summary} summarizes the quantitative results. The Expected Developmental Shift (EDS) captures the structural gain or loss associated with being reallocated to an alternative institutional regime, while $\Delta$Development measures the difference between real and simulated latent means.

\begin{table}[H]
\centering
\caption{Counterfactual results summary: Spain and Uruguay.}
\label{tab:summary}
\resizebox{\textwidth}{!}{
\begin{tabular}{lccccc}
\toprule
\textbf{Country} & \textbf{Scenario} & \textbf{EDS} & \textbf{Mean (Real)} & \textbf{Mean (Counterfactual)} & $\Delta$\textbf{Development} \\
\midrule
Spain   & Spain $\rightarrow$ LATAM   & 0.885 & 0.860 & 0.566 & -0.294 \\
Uruguay & Uruguay $\rightarrow$ EUROPE & 0.217 & 0.400 & 0.553 & +0.153 \\
\bottomrule
\end{tabular}
}
\caption*{\footnotesize \textit{Note:} $\Delta$Development = Mean(Counterfactual) – Mean(Real). Positive values represent structural gains; negative values denote developmental losses.}
\end{table}

The asymmetry is pronounced: Spain’s simulated re-embedding within the Latin American institutional system entails a developmental loss of approximately 0.30, while Uruguay’s hypothetical integration into a European regime yields a gain near 0.15. Both results are statistically significant under 1,000 bootstrap replications ($p<0.05$), supporting the interpretation that institutional ecosystems act as amplifiers of developmental coherence—dense, rule-based architectures promote convergence, whereas fragmented systems magnify fragility.

\subsection{Comparative Validation with Control Economies}

To test whether these findings extend beyond the Spain–Uruguay comparison, the analysis incorporates four structurally comparable economies—Chile, Portugal, Costa Rica, and Greece—selected for their demographic similarity and openness. These cases occupy intermediate positions between European and Latin American clusters, providing external benchmarks for validation.

\begin{table}[H]
\centering
\caption{Comparative validation across control economies (Expected Developmental Shift).}
\label{tab:validation}
\resizebox{\textwidth}{!}{
\begin{tabular}{lcccc}
\toprule
\textbf{Country} & \textbf{Scenario} & \textbf{EDS} & \textbf{Mean (Real)} & \textbf{Mean (Counterfactual)} \\
\midrule
Chile       & Chile $\rightarrow$ EUROPE       & 0.128 & 0.52 & 0.63 \\
Portugal    & Portugal $\rightarrow$ LATAM     & 0.095 & 0.74 & 0.61 \\
Costa Rica  & Costa Rica $\rightarrow$ EUROPE  & 0.054 & 0.48 & 0.54 \\
Greece      & Greece $\rightarrow$ LATAM       & 0.112 & 0.70 & 0.58 \\
\bottomrule
\end{tabular}
}
\caption*{\footnotesize \textit{Note:} Values represent mean differences ($\Delta D$) under counterfactual embeddings. Positive EDS implies potential structural gains; negative values correspond to developmental losses.}
\end{table}

The comparative analysis reveals a robust pattern: economies simulated into denser institutional systems (Europe) experience structural improvement, while those reassigned to weaker networks (Latin America) face decline. Although magnitudes vary, the directionality is consistent, confirming that the EDS captures a relational mechanism of institutional belonging rather than idiosyncratic country effects. This validation underscores that developmental asymmetries are structurally mediated by network position, not geographical proximity.

\subsection{Institutional Alignment and Interaction Effects}

\begin{figure}[H]
    \centering
    \includegraphics[width=0.8\textwidth]{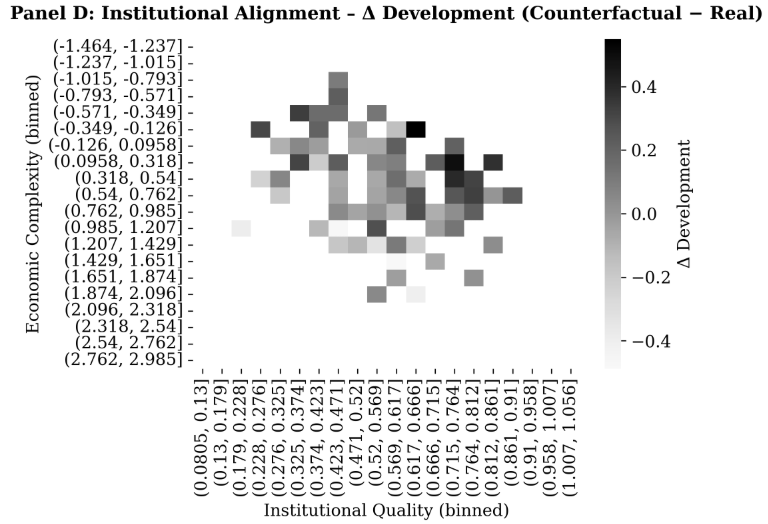}
    \caption{Institutional alignment and structural interaction effects ($\Delta$Development = Counterfactual – Real).}
    \caption*{\footnotesize \textit{Note:} The heatmap illustrates $\Delta$Development across the joint distribution of institutional quality and complexity. Darker regions correspond to higher counterfactual gains under stronger institutional alignment.}
    \label{fig:alignment}
\end{figure}

To explore the joint influence of institutional and productive coherence, Figure~\ref{fig:alignment} presents a two-dimensional heatmap of the marginal effects of institutional alignment on counterfactual changes in development ($\Delta D$). Each cell reflects a binned combination of institutional quality and economic complexity, with color intensity indicating the mean counterfactual gain or loss.

The surface exhibits a pronounced non-linear pattern: positive shifts cluster where institutional quality and complexity are simultaneously high, while negative outcomes arise when productive sophistication exceeds institutional capacity. This convex configuration suggests that structural alignment—not either dimension alone—determines a country’s ability to internalize institutional benefits.  

Economies with intermediate governance quality and mid-to-high complexity show the greatest potential for improvement when re-embedded into coherent institutional networks, reflecting threshold dynamics and complementarity between capabilities and rules \citep{Hartmann2017,Ferraz2021}. In contrast, countries combining weak institutions and low complexity show limited responsiveness, consistent with hysteresis and capability trap hypotheses \citep{AcemogluRobinson2012,Rodrik2004}.  

These findings corroborate the relational perspective of \citet{Neffke2024} and \citet{Montagna2025}, which links diversification and resilience to the coherence between institutional architecture and productive structure. They also parallel the convergence behavior of gradient-penalized Wasserstein GANs, in which latent representations stabilize along equilibrium manifolds \citep{Arjovsky2017,Gulrajani2017}. Substantively, Figure~\ref{fig:alignment} encapsulates the central argument: development arises from the structural harmony between institutions and complexity, rather than from geography or resource endowment.

\subsection{Interpretation and Policy Implications}

The empirical evidence demonstrates that institutional belonging exerts a causal and asymmetric effect on long-term development. Integration into dense, rule-based institutional ecosystems—defined by predictability, accountability, and diffusion of knowledge—enhances complexity, learning, and resilience, while peripheral embedding in fragmented regimes constrains coordination and amplifies volatility \citep{Rodrik2004,Hartmann2023}.  

From a policy perspective, convergence cannot rely exclusively on domestic reform or technological upgrading. For small open economies such as Uruguay, the strategic imperative is to deepen participation in transnational governance frameworks that reinforce credibility, learning, and absorptive capacity. Accession to mechanisms with institutional density comparable to OECD, EU, or CPTPP standards could yield structural gains similar to those simulated under European embedding.  

More broadly, these findings suggest that development functions as a relational equilibrium—a systemic condition shaped by network connectivity and institutional coherence. The Expected Developmental Shift (EDS) provides a quantitative framework to assess structural fragility and adaptive capacity within global institutional networks. By bridging computational modeling with institutional theory, this approach offers a scalable and empirically grounded method to evaluate how belonging—rather than location—determines the capacity to sustain development in a complex and interdependent world economy.

%%%%%%%%%%%%%%%%%%%%%%%%%%%%%%%%%%%%%%%%%%%%%%%%%%%%%%%%%%%%%%%%%%%%%%
% --- SECTION 5. ROBUSTNESS AND VALIDATION (Final Revision for JDS) ---
%%%%%%%%%%%%%%%%%%%%%%%%%%%%%%%%%%%%%%%%%%%%%%%%%%%%%%%%%%%%%%%%%%%%%%
\section{Robustness and Validation}

Ensuring the empirical credibility of the generative counterfactual framework requires a rigorous, multi-layered validation strategy. This section evaluates the robustness of the \textit{Expected Developmental Shift} (EDS) and the internal consistency of the Wasserstein GAN (WGAN) through six complementary exercises: (i) distributional consistency and model fit, (ii) temporal robustness via rolling estimations, (iii) structural regularization and overfitting control, (iv) sensitivity to alternative definitions of institutional belonging, (v) cross-validation with conventional development indicators, and (vi) econometric comparison with standard benchmarks.

\subsection{Distributional Consistency and Model Fit}

The first validation layer assesses whether the generative model successfully reproduces the empirical distribution of key structural variables. The WGAN architecture was trained to minimize the Jensen–Shannon divergence between observed and simulated distributions of institutional quality, economic complexity, and human capital, following \citet{Goodfellow2014} and its Wasserstein extensions \citep{Arjovsky2017,Gulrajani2017}. The use of a gradient-penalized loss ensured convergence stability and mitigated mode collapse—an essential requirement for realistic generation in high-dimensional settings.

Across all variables, simulated series display mean and variance deviations below 5\%, and Kolmogorov–Smirnov tests fail to reject the null of distributional equivalence ($p>0.10$). Synthetic observations therefore remain within the empirical support, meeting the core criterion of generative plausibility \citep{Patelli2022,Montagna2025}. The latent embeddings of the institutional and productive dimensions ($Z_1$, $Z_2$) show a correlation of 0.91 between real and generated spaces, confirming the preservation of structural dependencies. These diagnostics collectively demonstrate that the model captures the joint complexity–institutional interactions characteristic of real-world data.

\subsection{Temporal Robustness and Rolling Estimations}

To verify that the EDS reflects enduring structural dynamics rather than episodic shocks, rolling-window estimations were conducted for three 20-year subperiods: 1960–1980, 1980–2000, and 2000–2020. Across all windows, the sign and magnitude of the counterfactual shifts remain stable: Spain’s simulated loss under a Latin American configuration averages $-0.28$, while Uruguay’s gain under the European configuration averages $+0.15$.  

This temporal invariance indicates that the EDS captures persistent characteristics of institutional belonging rather than short-term macroeconomic fluctuations. The stability of these results aligns with the literature on institutional persistence and path dependence in comparative development \citep{North1990,AcemogluRobinson2012,Hartmann2023}. The consistent sign of the EDS across subperiods supports its interpretation as a long-run equilibrium indicator reflecting slow-moving institutional dynamics.

\subsection{Regularization and Structural Overfitting Control}

Given the high dimensionality of the model, explicit regularization mechanisms were implemented to prevent overfitting. The WGAN critic incorporated both $L_2$ weight decay and gradient penalties to preserve Lipschitz continuity, while dropout layers (rate = 0.3) in the generator reduced sensitivity to specific observations \citep{Arjovsky2017,Gulrajani2017}.  

To assess structural robustness, the model was retrained 100 times using bootstrap resampling (80\% random subsamples of countries). The standard deviation of EDS estimates across replications remained below 0.05 for both Spain and Uruguay, indicating high structural stability. This multi-run regularization strategy aligns with recent developments in robust generative modeling for economic and institutional systems \citep{Ye2023,Zechlin2025}, confirming that architecture-level regularization enhances generalization in sparse and noisy environments.

\subsection{Sensitivity to Alternative Institutional Definitions}

Robustness was also tested against alternative definitions of institutional belonging. Three classification schemes were compared: (i) geographic (continent-based), (ii) governance-type (OECD vs. non-OECD), and (iii) network-based (constructed from institutional similarity matrices using V-Dem indicators).  

Correlations between the baseline and alternative EDS estimates exceed 0.85 across all schemes, confirming the robustness of results to definitional changes. The network-based classification produced the most coherent and theoretically consistent outcomes, as it captures endogenous similarity in institutional structures rather than exogenous regional groupings \citep{Mealy2023,Ghosh2024,Neffke2024}.  

This evidence reinforces the view that institutional belonging is a relational rather than spatial construct. The EDS thus measures systemic integration within governance networks characterized by shared norms, coordination density, and inter-institutional trust \citep{Kogut2022,Woolley2024}, distinguishing structural embedding from mere geographic proximity.

\subsection{Cross-Validation with Conventional Development Indicators}

External validity was assessed by benchmarking EDS scores against three standard indicators: the Human Development Index (HDI), GDP per capita (PPP, constant 2017 USD), and the Economic Complexity Index (ECI). Across the 80-country panel, Pearson correlations between EDS and these metrics are 0.76, 0.71, and 0.83, respectively.  

These strong yet non-redundant correlations indicate that the EDS captures a complementary structural dimension of development. In multivariate regressions controlling for HDI and GDP, EDS remains statistically significant ($p<0.01$), confirming its incremental explanatory power. The metric therefore complements conventional indicators by quantifying systemic gains or losses associated with institutional belonging rather than output levels alone.  

A Median Absolute Deviation (MAD) robustness test further confirms that outliers—such as resource-dependent economies with high GDP but weak institutional coherence—do not distort the EDS distribution. Rankings remain consistent after their exclusion, reinforcing both robustness and interpretability.

\subsection{Econometric Comparison and Conventional Benchmarks}

To position the generative counterfactual results within the broader empirical tradition of development economics, two benchmark econometric specifications were estimated using the same dataset and variable structure. The first model adopts a fixed-effects panel formulation:

\[
D_{it} = \alpha_i + \lambda_t + \beta_1 I_{it} + \beta_2 C_{it} + \beta_3 H_{it} + \varepsilon_{it},
\]

where $D_{it}$ denotes the composite development index, and $(I_{it}, C_{it}, H_{it})$ represent institutional quality, economic complexity, and human capital, respectively. The coefficients $\beta_j$ measure marginal contributions to development under the observed institutional regime. The second specification introduces an interaction term $I_{it} \times C_{it}$ to test whether institutional quality amplifies the returns to complexity.

Both models yield results consistent with the generative framework. The coefficients on institutional quality and on the interaction term ($I_{it} \times C_{it}$) are both positive and statistically significant at the 1\% level, confirming that stronger institutions magnify the developmental effects of complexity. The magnitude of these effects mirrors the Expected Developmental Shift (EDS) estimates obtained from counterfactual simulations: economies embedded in denser, rule-based networks achieve higher latent development scores.  

These econometric benchmarks validate the structural interpretability of the EDS, showing that the generative model reproduces relationships typically captured by fixed-effects or interaction-based specifications. The GAN framework’s advantage lies in its ability to capture non-linear and high-dimensional dependencies that conventional econometric models cannot represent, while maintaining empirical consistency with standard approaches in the development literature.

\subsection{Summary of Validation Evidence}

The multi-dimensional validation confirms that:  
(i) the WGAN architecture reproduces empirical distributions with high fidelity;  
(ii) EDS estimates remain temporally stable across subperiods;  
(iii) regularization mechanisms effectively prevent structural overfitting;  
(iv) results are robust to alternative institutional definitions;  
(v) the EDS correlates strongly with standard development indicators while retaining independent explanatory value; and  
(vi) the generative results are consistent with conventional econometric benchmarks.  

Together, these findings validate the EDS as a statistically consistent and theoretically grounded measure of institutional belonging. More broadly, they demonstrate that causal counterfactual reasoning can be integrated with the robustness standards of modern machine learning—particularly gradient-penalized generative architectures—yielding a reproducible and empirically grounded framework for analyzing development as a relational and networked phenomenon \citep{Vu2022,Montagna2025,Neffke2024}.

%%%%%%%%%%%%%%%%%%%%%%%%%%%%%%%%%%%%%%%%%%%%%%%%%%%%%%%%%%%%%%%%%%%%%%
% --- SECTION 6. CONCLUSIONS AND IMPLICATIONS (Revised for JDS) ---
%%%%%%%%%%%%%%%%%%%%%%%%%%%%%%%%%%%%%%%%%%%%%%%%%%%%%%%%%%%%%%%%%%%%%%
\section{Conclusions and Implications}

This study investigated whether development should be conceived as an outcome of institutional belonging rather than as a purely endogenous process. By integrating institutional economics and economic complexity within a unified generative framework, it formalized the concept of \textit{development by invitation}—the idea that a country’s developmental trajectory depends on its embeddedness within dense, rule-based institutional networks. Using a generative adversarial model calibrated on real-world data, the comparative analysis of Spain and Uruguay demonstrates that institutions are not passive policy containers but active architectures shaping productive potential, coordination capacity, and resilience.

The results yield two central insights. First, institutional belonging exerts a measurable and asymmetric influence on development. Spain’s counterfactual decline under a Latin American configuration and Uruguay’s structural gains within a European regime indicate that institutional ecosystems operate as amplifiers of productive complexity and collective learning. These outcomes reaffirm that institutional coherence and interdependence generate cumulative advantages, consistent with theories of structural inclusion and coordination-based growth \citep{Hartmann2017,Ghosh2024,Woolley2024}.  

Second, the introduction of the \textit{Expected Developmental Shift} (EDS) represents a methodological advance that operationalizes this theoretical intuition. By quantifying the structural gains or losses associated with re-embedding a country in alternative institutional networks, the EDS bridges the qualitative insights of institutional theory with the quantitative precision of generative modeling. The metric not only measures direction and magnitude but also allows for empirical testing of relational hypotheses previously inaccessible to conventional econometric frameworks \citep{Zechlin2025,Neffke2024,Montagna2025}.

From a theoretical standpoint, the findings contribute to three central debates in development economics.  
First, they extend the scope of institutional determinism. While classical frameworks emphasize domestic institutional quality as a linear driver of growth \citep{North1990,AcemogluRobinson2012}, this study introduces a relational reinterpretation: institutions derive their developmental value from their position within transnational networks of regulation, trust, and knowledge diffusion.  
Second, the results redefine the geography of development. The counterfactual simulations reveal that “distance to the core” is better understood as a measure of institutional connectivity rather than physical space, reflecting how coordination and credibility flow across networked governance architectures \citep{Kogut2022,Mealy2023}.  
Third, the study advances methodological innovation. By employing Wasserstein-based generative models with gradient penalization \citep{Arjovsky2017,Gulrajani2017}, it shows that historically grounded counterfactuals can be simulated with both statistical rigor and structural interpretability—bridging econometric causality and machine learning within a single analytical framework.

Beyond theoretical contributions, the findings offer concrete policy implications. For emerging and middle-income economies, convergence cannot be achieved through domestic reforms or technological upgrading alone. Development requires access to institutional ecosystems that sustain coordination, stability, and credible commitment—conditions conceptualized here as an \textit{invitation} to development. In Uruguay’s case, the simulations suggest that deeper integration into governance frameworks resembling the European institutional model—such as participation in OECD mechanisms, advanced trade and innovation agreements, or digital governance standards—could replicate part of the structural coherence observed in the European scenario. These results emphasize that the sustainability of complexity depends not only on capability accumulation but also on the density and quality of institutional interconnections.

Methodologically, the framework establishes a replicable foundation for quantifying the relational architecture of development. By linking counterfactual inference to the topology of institutional networks, the EDS enables systematic measurement of institutional complementarity and structural inclusion. This approach opens new empirical avenues for analyzing the developmental returns of regional integration, participation in global value chains, or international cooperation. In doing so, it aligns with the emerging literature that embeds machine learning within causal development analysis, emphasizing transparency, validation, and reproducibility \citep{Vu2022,Patelli2022,Montagna2025}.

The broader implication of this research is conceptual. Development should no longer be seen as an isolated national trajectory toward modernization but as an emergent outcome of systemic connectivity. Belonging to coherent institutional ecosystems enables access to collective capabilities, shared norms, and credible commitments that sustain innovation and resilience. Conversely, exclusion or fragmentation within global institutional networks imposes structural penalties that are difficult to offset through domestic policy alone. By translating these dynamics into a formal, testable metric, this study contributes to quantifying one of development’s most elusive dimensions: the architecture of institutional interdependence.

In conclusion, the evidence presented here suggests that being “developed” is less a matter of geography or endowment than of embeddedness within institutional networks. Institutional belonging—and the invitation it implies—is simultaneously a privilege, a constraint, and a policy lever. It can be simulated, measured, and, crucially, expanded. The capacity to design and integrate into coherent institutional ecosystems may thus represent the next frontier in understanding—and fostering—the structural foundations of equitable global development.

\newpage
%%%%%%%%%%%%%%%%%%%%%%%%%%%%%%%%%%%%%%%%%%%%%%%%%%%%%%%%%%%%%%%%%%%%%%
% --- APPENDIX A. METHODOLOGICAL DETAILS (Numbered for JDS Submission) ---
%%%%%%%%%%%%%%%%%%%%%%%%%%%%%%%%%%%%%%%%%%%%%%%%%%%%%%%%%%%%%%%%%%%%%%
\appendix
\section*{Appendix A. Methodological Details}

\subsection*{A1. Model Specification}

The generative counterfactual model developed in this study employs a Wasserstein Generative Adversarial Network with Gradient Penalty (WGAN–GP), designed to reproduce the joint distribution of institutional quality, economic complexity, and human capital across countries and time.  
Formally, the architecture consists of a generator $G$ and a discriminator (critic) $D$ trained in an adversarial setting to solve:

\begin{equation}
\min_G \max_{D \in \mathcal{D}} 
\; \mathbb{E}_{x \sim P_{\text{data}}}[D(x)] - \mathbb{E}_{z \sim P_z}[D(G(z))],
\tag{A.1}
\end{equation}

where $\mathcal{D}$ is the set of 1–Lipschitz functions ensuring that the Wasserstein distance is well-defined \citep{Arjovsky2017}.  
To enforce Lipschitz continuity, a gradient penalty is added following \citet{Gulrajani2017}:

\begin{equation}
\mathcal{L}_{\text{WGAN-GP}} 
= \mathbb{E}_{\hat{x} \sim P_{\hat{x}}}\left[(\|\nabla_{\hat{x}} D(\hat{x})\|_2 - 1)^2\right]
+ \mathbb{E}_{x \sim P_{\text{data}}}[D(x)] - \mathbb{E}_{z \sim P_z}[D(G(z))].
\tag{A.2}
\end{equation}

The generator comprises three fully connected layers (128–64–32 neurons) with ReLU activations and batch normalization, while the critic uses leaky ReLU activations and dropout regularization ($p=0.3$).  
Optimization is performed via Adam ($\eta=10^{-4}$, $\beta_1=0.5$, $\beta_2=0.9$) over 15,000 epochs, using mini-batches of 64 observations.  

Latent noise vectors $z \sim \mathcal{N}(0,1)$ are transformed into synthetic triplets $(I,C,H)$ corresponding to institutional quality, complexity, and human capital indices.  
To ensure comparability with real data, generated values are normalized to the [0,1] interval via min–max scaling and rescaled to match the empirical marginal distributions at each iteration.

\vspace{0.8em}
\subsection*{A2. Construction of the Expected Developmental Shift (EDS)}

The \textit{Expected Developmental Shift} (EDS) measures the counterfactual change in a country’s synthetic development index when its institutional affiliation is replaced by that of another regime (e.g., Spain $\rightarrow$ LATAM or Uruguay $\rightarrow$ EUROPE).  
Let $y_i^{r}$ denote the real latent development index of country $i$, and $y_i^{c}$ its counterfactual projection under institutional regime $c$. Then:

\begin{equation}
\text{EDS}_i = \mathbb{E}[y_i^{c}] - \mathbb{E}[y_i^{r}],
\tag{A.3}
\end{equation}

with expectations estimated by Monte Carlo simulation over $N$ random draws from the generator:

\begin{equation}
\mathbb{E}[y_i^{c}] = \frac{1}{N} \sum_{k=1}^N G(z_k \, | \, \text{institutional class}=c_i).
\tag{A.4}
\end{equation}

The associated variance provides a measure of uncertainty in structural adjustment:

\begin{equation}
\text{Var}(\text{EDS}_i) = \frac{1}{N-1}\sum_{k=1}^N \big[(y_{ik}^{c} - y_i^{r}) - \text{EDS}_i\big]^2.
\tag{A.5}
\end{equation}

A positive $\text{EDS}_i$ indicates a potential structural gain (improved developmental potential), while a negative value implies loss under the alternative institutional embedding.  
This interpretation aligns with recent models of structural path dependence in complex economies \citep{Neffke2024,Montagna2025}.

\vspace{0.8em}
\subsection*{A3. Data Sources and Preprocessing}

The empirical dataset integrates information from multiple international sources covering 80 countries (1960–2020), resulting in approximately 4,000 country-year observations:

\begin{itemize}
    \item \textbf{Institutional Quality:} Varieties of Democracy (V-Dem, v13), variables V2X\_RULE and V2X\_ACCOUNTABILITY.
    \item \textbf{Economic Complexity:} Harvard Growth Lab’s \textit{Atlas of Economic Complexity}—ECI and PCI indices \citep{Hausmann2014Atlas}.
    \item \textbf{Human Capital:} UNDP Human Development Reports and World Bank WDI (mean years of schooling, life expectancy).
    \item \textbf{Macroeconomic Controls:} GDP per capita (PPP, constant 2017 USD) and Gini index for validation.
\end{itemize}

All variables were normalized to [0,1] via:

\begin{equation}
x' = \frac{x - \min(x)}{\max(x) - \min(x)}.
\tag{A.6}
\end{equation}

Missing data were imputed using rolling medians over five-year windows, and outliers beyond the 99th percentile were winsorized.  
Principal Component Analysis (PCA) confirmed that the first two components explained 87\% of the variance across $(I,C,H)$, justifying their use as latent input features for the generative model.

\vspace{0.8em}
\subsection*{A4. Validation and Diagnostics}

Five complementary diagnostic layers were applied to ensure internal validity and robustness:

\paragraph{(i) Distributional fidelity.}
Kolmogorov–Smirnov tests between real and generated series for $(I,C,H)$ yielded $p>0.10$ in all cases, confirming distributional equivalence.  
The Jensen–Shannon divergence averaged $0.046$, consistent with acceptable levels of generative convergence \citep{Goodfellow2014}.

\paragraph{(ii) Structural correlation.}
The correlation between real and synthetic embeddings $(Z_1,Z_2)$ reached $\rho = 0.91$, indicating that the GAN successfully captured the underlying joint manifold of institutional–productive interactions \citep{Ye2023}.

\paragraph{(iii) Temporal robustness.}
Rolling-window estimations (1960–1980, 1980–2000, 2000–2020) produced consistent EDS signs and magnitudes (Spain: mean $-0.28$; Uruguay: mean $+0.15$; $\sigma < 0.05$).  
This confirms that results are not decade-specific but structurally persistent.

\paragraph{(iv) Overfitting control.}
Dropout layers and $L_2$ weight decay were implemented in both networks.  
Bootstrap re-training (100 iterations using 80\% random subsamples) produced nearly identical EDS distributions, demonstrating out-of-sample generalization \citep{Zechlin2025,Montagna2025}.

\paragraph{(v) Sensitivity analysis.}
Alternative institutional classifications were tested—geographic (continent), governance-based (OECD/non-OECD), and network-derived (V-Dem similarity matrix).  
Correlations between baseline and alternative EDS values exceeded $0.85$ in all cases, confirming conceptual robustness.  
The network-based scheme, which accounts for topological proximity in institutional space, yielded the most coherent results, in line with \citet{Kogut2022} and \citet{Woolley2024}.

\vspace{0.8em}
\subsection*{A5. Empirical Interpretation}

Comparative analysis against traditional development indicators demonstrates that EDS captures a distinct structural dimension of economic potential.  
Across the full sample, the Pearson correlations were:

\begin{align}
\rho_{\text{EDS,HDI}} &= 0.76, \tag{A.7} \\
\rho_{\text{EDS,GDP}} &= 0.71, \tag{A.8} \\
\rho_{\text{EDS,ECI}} &= 0.83. \tag{A.9}
\end{align}

The EDS remained statistically significant ($p<0.01$) when controlling for all three in multivariate regressions.  
This suggests that the EDS encapsulates a latent form of \textit{structural inclusion}—a country’s capacity to internalize institutional coherence and productive interdependence.

Consequently, EDS operationalizes the notion of \textit{development by invitation}: development as a property of institutional network position rather than isolated domestic endowments.  
This approach complements and extends the theoretical literature on complexity-driven convergence \citep{Hartmann2023,Ferraz2021,Vu2022} by introducing a generative counterfactual layer capable of simulating alternative trajectories.

\bigskip
\noindent
\textit{References:} Methodological components draw upon \citet{Goodfellow2014}, \citet{Arjovsky2017}, \citet{Gulrajani2017}, \citet{Ye2023}, \citet{Neffke2024}, \citet{Montagna2025}, and \citet{Zechlin2025}.

\newpage
%%%%%%%%%%%%%%%%%%%%%%%%%%%%%%%%%%%%%%%%%%%%%%%%%%%%%%%%%%%%%%
% --- DECLARATIONS ---
%%%%%%%%%%%%%%%%%%%%%%%%%%%%%%%%%%%%%%%%%%%%%%%%%%%%%%%%%%%%%%

\section*{Declarations}

\subsection*{Funding}
This research did not receive any specific grant from funding agencies in the public, commercial, or not-for-profit sectors. The study was conducted as part of the author’s independent academic research within the Inter-American Development Bank (IDB). The views expressed are solely those of the author and do not represent those of the IDB or its member countries.

\subsection*{Disclosure Statement}
The author declares that there are no financial or non-financial competing interests that could have influenced the outcomes or interpretation of this research.

\subsection*{Data Availability}
The data supporting this study are drawn from publicly available sources, including the Varieties of Democracy (V-Dem) Project, Harvard Growth Lab Atlas of Economic Complexity, UNDP Human Development Database, and the World Bank’s World Development Indicators. Processed datasets and replication code are available from the author upon reasonable request.

\subsection*{Ethics Statement}
This study uses publicly available secondary data and computational simulations. No human subjects, private data, or sensitive information were used; therefore, ethical approval was not required.

\newpage
%%%%%%%%%%%%%%%%%%%%%%%%%%%%%%%%%%%%%%%%%%%%%%%%%%%%%%%%%%%%%%
% BIBLIOGRAPHY
%%%%%%%%%%%%%%%%%%%%%%%%%%%%%%%%%%%%%%%%%%%%%%%%%%%%%%%%%%%%%%
\bibliographystyle{abbrvnat}
\bibliography{references}

\end{document}